\documentclass[epj,nopacs]{svjour}

\usepackage{latexsym}
\usepackage[numbers,sort&compress]{natbib}
\usepackage{graphics}
\usepackage{graphicx}
\usepackage{subfigure}
\usepackage{hyperref}
\usepackage[tbtags]{amsmath}
\usepackage{color}
\usepackage{amsmath}
\usepackage{stfloats}

\begin{document}

\title{Birkhoff's Theorem in $f(T)$ Gravity up to the Perturbative Order}
\author{Han Dong \inst{1} \thanks{\email{donghan@mail.nankai.edu.cn}},
Ying-bin Wang \inst{1} \thanks{\email{wybn@mail.nankai.edu.cn}}
\and Xin-he Meng\inst{1,2} \thanks{\email{xhm@nankai.edu.cn}}
}

\institute{Department of physics, Nankai University, Tianjin 300071, China
\and Kavli Institute of Theoretical Physics China, CAS, Beijing 100190, China}

\date{Received: date / Revised version: date}

\abstract{$f(T)$ gravity, a generally modified teleparallel gravity,
has become very popular in recent times as it is able to reproduce
the unification of inflation and late-time acceleration
without the need of a dark energy component or an inflation field.
In this present work, we investigate specifically the range of validity of Birkhoff's theorem
with the general tetrad field via perturbative approach. At zero order, Birkhoff's theorem is valid
and the solution is the well known Schwarzschild-(A)dS metric.
Then considering the special case of the diagonal tetrad field, we present a new spherically symmetric solution
in the frame of $f(T)$ gravity up to the perturbative order.
The results with the diagonal tetrad field satisfy the physical equivalence
between the Jordan and the so-called Einstein frames, which are realized via conformal transformation,
at least up to the first perturbative order.
\\\\
PACS.98.80.Cq Modified theories of gravity }
\maketitle

\section{Introduction}

Since the discovery of the accelerating expansion of the universe evolution,
people have made great efforts to investigate the hidden mechanism,
which also provides us with great opportunities to deeply
probe the fundamental theories of gravity dominating the cosmic evolution.
Recently, a new modified gravity to account for the accelerating expansion of the universe,
i.e., $f(T)$ gravity, has been proposed by extending the action of teleparallel gravity to a general term,
which is built on teleparallel geometry.
The framework is a generalization of the so-called \textit{Teleparallel Equivalent of
General Relativity} (TEGR) which was firstly propounded by Einstein in
1928 \cite{einstein1,einstein2} to unify gravity and electromagnetism,
and then was revived as a geometrical alternative to the
Riemannian geometry of general relativity in the 1960s
(for some reviews, see \cite{TEGR1,TEGR2}).
Contrary to the theory of general relativity, which is based on
Riemann geometry involving only curvature,
the TEGR is based on the so called Weitzenb\"ock geometry,
with the non-vanishing torsion. Owing to the definition of
Weitzenb\"ock connection rather than the Levi-Civita connection for the Riemann geometry,
the Riemann curvature is automatically vanishing in the TEGR framework,
which brings the theory a new name, \textit{Teleparallel Gravity}.
For a specific choice of parameters, the TEGR behaves completely
equivalent to Einstein's theory of general relativity.

Similar to the generalization of Einstein's theory of general relativity to
$f(R)$ gravity and other modified gravity theories (for some references,
see \cite{fR0,fRrev,fR1,fR11,fRa0,fRa1,fRa2,fRa3,fRa4,fRb1,fRb2,fRb3,fRb4,fRb5,fRb6,fRa6,fRa7,fRa8,fRa9,fR2,fR3,fR4,fR5,fR6,fR7,fR8,fR9}),
the modified version of teleparallel gravity assumes a general function
$f(T)$ of the torsion $T$ as the model Lagrangian density.
Also, $f(T)$ gravity can be directly reduced to the TEGR
if we choose the simplest case, that is, $f(T)\!\!=\!\!T$.
As one of modified gravitational theories, $f(T)$ gravity
is firstly invoked to drive inflation by Ferraro and Fiorini \cite{fT0}.
Later, Bengochea and Ferraro \cite{fT}, as well as Linder \cite{fT1},
propose to use $f(T)$ theory to drive the current accelerated expansion
of our universe without introducing the mysterious dark energy component.
The Lorentz invariance \cite{fT_Lorentz} and conformal invariance \cite{fT_conformal}
of $f(T)$ gravity are also investigated, besides many interesting results presented.
In our previous works \cite{fT_birkhoff1,fT_birkhoff2},
we investigated the validity of Birkhoff's theorem in $f(T)$ gravity,
also discussed the equivalence between both the Einstein frame and Jordan frame.
Furthermore, by using the general function $f(T)$ of torsion scalar as the Lagrangian density,
$f(T)$ gravity can deduce a field equation with second order only,
instead of the fourth order as in the Einstein-like field equation of the general $f(R)$ gravity,
and avoids the instability problems caused by higher-order derivatives
as demonstrated in the metric framework $f(R)$ gravity models.
This feature has led to a rapidly increasing interest
to explore various aspects of $f(T)$ gravity models in the literature.
Some new $f(T)$ forms are proposed \cite{fT2,fT6}.
Some people give constraints on $f(T)$ gravity \cite{fT7,fT3} by using the latest observational data,
analyzing the dynamical behavior \cite{fT4} and the cosmic large scale structure \cite{fT_LSS1,fT_LSS2},
studying the relativistic neutron star \cite{fT_relativistic_stars},
the matter bounce \cite{fT_bounce} and the perturbations \cite{pt1,pt2,pt3} in $f(T)$ gravity framework.
investigating the static spherical symmetry solutions,
the equation-of-state parameter crossing the phantom divide \cite{fT_w1,fT_w2}.
Some other relevant work can be seen in references \cite{fT5,fT8,fT9,fT10} and in a newly review \cite{fT11}.

Birkhoff's theorem is also called the Jebsen-Birkhoff theorem,
for it was actually discovered by Jebsen, two years
before George D. Birkhoff in 1923 \cite{birkhoff,bb1,bb2,bb3}.
The theorem states that the spherically
symmetric gravitational field in vacuum must be static,
with a metric uniquely given by the Schwarzschild solution
form of Einstein equations \cite{weinberg}.
It is well known that the Schwarzschild metric is found in 1918 as
the external (vacuum) solution of a static and spherical symmetric star.
Birkhoff's theorem means that any spherically
symmetric object possesses the same static gravitational field,
as if the mass of the object were concentrated at the center.
Even if the central spherical symmetric object is dynamic motion,
such as the case in the collapse and pulsation of stars,
the external gravitational field is still static
if only the radial motion is spherically symmetric.
The same feature occurs in classical Newtonian gravity,
also in some case of static electronic-magnetism analogously.

As we know, for reconstructing the action $f(R)$
containing of inflationary and dark energy epochs,
it is often easily done by introducing an auxiliary
scalar field via the conformal transformation,
because it is equivalent to a kind of Brans-Dicke theory
with a non-propagating scalar field and a non-null potential.
However, the equivalence of both approaches via the conformal transformation
seem to be valid prior to the order of perturbations,
where both theories seem to exhibit different behaviors.
At first linear order perturbation,
Birkhoff's theorem generally does not hold in the frame of $f(R)$ gravity by
using its scalar-tensor representation,
while strong restrictions are imposed on the scalar curvature
and on the scalar field, respectively for its validity \cite{invalid_fR3}.
The perturbative result in the Jordan frame is different from that in the Einstein frame,
which also indicates the different physical meaning between
Einstein frame and Jordan frame at least in a perturbative approach \cite{invalid_fR2,fR_perturbation}.
Differing from the case in $f(R)$ gravity,
there is an additional scalar-torsion coupling term
present in the action \cite{fT_conformal}.
Therefore, $f(T)$ gravity is not simply dynamically equivalent to
the TEGR action plus a scalar field via conformal transformation,
and one cannot apply the results of scalar-tensor theories directly to $f(T)$ gravity.
Beyond that, the field equation with second order only, deduced for $f(T)$ gravity,
makes more clear for a picture of the range of the validity of Birkhoff's theorem
and the physical equivalence between both frames in a perturbative approach.

In this work we investigate Birkhoff's theorem
in $f(T)$ gravity using a perturbative approach,
and compare the results in the so-called Einstein and Jordan frames.
The physical equivalence between both frames is discussed at least in perturbation order.
First, in section two we briefly review $f(T)$ theories,
and in section three we represent $f(T)$ gravity by conformal transformation.
Second, in section four we investigate the range of validity
of Birkhoff's theorem with the general tetrad in $f(T)$ gravity via perturbative approach.
Considering the special case of the diagonal tetrad,
we present a new spherically symmetric solution,
and both the Jordan and Einstein frames are discussed in this section.
Finally, we summarize some conclusions with discussion in the last section.

\section{Elements of $f(T)$ Gravity}

Instead of the metric tensor, the vierbein field $\mathbf{e}_{i}(x^{\mu})$,
who can compose into the metric tensor,
plays the role of the dynamical variable in the teleparallel gravity.
It is defined as the orthonormal basis of the tangent space
at each point $x^{\mu}$ in the manifold, namely,
$\mathbf{e}_{i}\cdot \mathbf{e}_{j}\!\!=\!\!\eta_{ij}$,
where $\eta_{ij}\!\!=\!\!diag(1,-1,-1,-1)$ is the Minkowski metric.
The vierbein vector can be expanded in space-time coordinate basis:
$\mathbf{e}_{i}\!\!=\!\!e^{\mu}_{i} \partial_{\mu}$,
$\mathbf{e}^i\!\!=\!\!e^i_\mu{\rm d}x^\mu$. According to the convention,
Latin indices and Greek indices, both running from 0 to 3,
label the tangent space coordinates and the space-time
coordinates, respectively. The components of vierbein are related by
$e_{\mu}^i e^{\mu}_j\!\!=\!\!\delta^{~i}_{j}$, ~~$e_{\mu}^i e^{\nu}_i\!\!=\!\!\delta_{\mu}^{~\nu}$.

The metric tensor is determined uniquely by the vierbein as
\begin{equation}\label{basic}
g_{\mu\nu}=\eta_{ij} e_{\mu}^i e_{\nu}^i,
\end{equation}
which can be equivalently expressed as $\eta_{ij}\!\!=\!\!g_{\mu\nu} e^i_{\mu} e^j_{\nu}$.
The definition of the torsion tensor is given then by
\begin{equation}
T^{\rho}_{~\mu\nu}=\Gamma^{\rho}_{~\nu\mu}-\Gamma^{\rho}_{~\mu\nu}.
\end{equation}
where $\Gamma^{\rho}_{~\mu\nu}$ is the connection.
Evidently, $T^{\rho}_{~\mu\nu}$ vanishes in the Riemann geometry
since the Levi-Civita connection is symmetric
with respect to the two covariant indices.
Differing from that in Einstein's theory of general relativity,
teleparallel gravity uses Weitzenb\"ock connection,
defined directly from the vierbein:
\begin{equation}\label{vierbein}
\Gamma^{\rho}_{~\mu\nu}=e_i^{\rho} \partial_{\nu} e^i_{\mu}.
\end{equation}
Accordingly, the antisymmetric non-vanishing torsion is
\begin{equation}\label{torsion}
T^{\rho}_{~\mu\nu}=e_i^{\rho}(\partial_{\mu}e^i_{\nu} - \partial_{\nu}e^i_{\mu}).
\end{equation}
It can be confirmed that the Riemann curvature in this framework is precisely vanishing:
\begin{equation}
R^\rho_{~\theta\mu\nu}=\partial_\mu \Gamma^\rho_{~\theta\nu}-\partial_\nu \Gamma^\rho_{~\theta\mu}+\Gamma^\rho_
{~\sigma\mu}\Gamma^\sigma_{~\theta\nu}-\Gamma^\rho_{~\sigma\nu} \Gamma^\sigma_{\theta\mu}=0.
\end{equation}

In order to get the action of the teleparallel gravity, it is
convenient to define other two tensors:
\begin{equation}\label{contorsion}
K^{\mu\nu}_{~~\rho}=-\frac{1}{2}(T^{\mu\nu}_{~~\rho}-T^{\nu\mu}_{~~\rho}-T_{\rho}^{~\mu\nu})
\end{equation}
and
\begin{equation}\label{S}
S_\rho^{~\mu\nu}=\frac{1}{2}(K^{\mu\nu}_{~~\rho}+\delta_\rho^{~\mu}T^{\theta\nu}_{~~\theta}-\delta_\rho^{~\nu}T^
{\theta\mu}_{~~\theta}).
\end{equation}
Then the torsion scalar as the teleparallel Lagrangian density is defined by
\begin{equation}\label{T}
T=S_{\rho}^{~\mu\nu} T^{\rho}_{~\mu\nu}.
\end{equation}
The action of teleparallel gravity is then expressed as
\begin{equation}
I=\frac{1}{16\pi G}\int {\rm d}^4 x~e\,T ,
\end{equation}
where $e\!\!=$det$(e^i_{\mu})\!\!=\!\!\sqrt{-g}$. Performing variation of the
action with respect to the vierbein, one can directly get the equations of
motion which are equivalent to the results of Einstein's theory of
general relativity in some sense.

Just as in $f(R)$ theory, the generalized version of
teleparallel gravity could be obtained by extending the Lagrangian
density directly to a general function of the scalar torsion $T$:
\begin{equation}\label{action}
 I=\frac{1}{16\pi G}\int {\rm d}^4x~e\,f(T).
\end{equation}
This modification is expected possibly to provide a natural way
to understand the cosmological observations,
especially for the dark energy phenomena, as a motivation.
Then the variation of the action with respect to vierbein,
which is posted in the Appendix, leads to the following equations:
\begin{equation}\label{field eqn}
\begin{split}
{\big[}e^{-1}e^i_\mu\partial_\sigma(eS_i^{~\sigma\nu})-T^\rho_{~\sigma\mu}S_\rho^{~\nu\sigma}{\big]}f_T+
S_\mu^{~\rho\nu}\partial_\rho T f_{TT}\\
-\frac{1}{4}\delta_\mu^{~\nu}f=4\pi GT_\mu^{~\nu} ,
\end{split}
\end{equation}
where $f_T$ and $f_{TT}$ represent the first and
second-order derivatives with respect to $T$, respectively,
and $S_i^{~\sigma\nu}\!\!=\!\!e_i^\rho S_\rho^{~\sigma\nu}$.
~$T_\mu^{~\nu}$ is the energy-momentum tensor of the particular matter,
assuming that matter couples to the metric in the standard form.

\section{Represent the $f(T)$ Gravity by Conformal Transformation}

It is well known that $f(R)$ gravity is dynamically equivalent
to a particular class of scalar-tensor theories via conformal transformation,
while Birkhoff's theorem generally does not hold in scalar-tensor gravity.
The case of $f(T)$ gravity via conformal transformation is more complicated
than that of $f(R)$ theories, which has been proved in the work \cite{fT_conformal}.
In this section, we explore the difference between $f(T)$ gravity and scaler-tensor theory,
and compare the results obtained, respectively from the Jordan and Einstein frames via conformal transformation.
Firstly, the general action for a Brans-Dicke-like $f(T)$ theory can be write in the Jordan frame as,
\begin{equation}\label{SBD}
    S_{BD} = \int {\rm d}^4 x~e\bigg[ \phi T - \frac{\omega}{\phi}g^{\mu\nu} \nabla_{\mu}\phi \nabla_{\nu}\phi
    - V(\phi) + 2k^2\mathcal{L}_m(e_{\mu}^{~i}) \bigg],
\end{equation}
where we assume $\omega$ to be constant.

By the conformal transformation, we can get the tilded tetrad and metric of Einstein frame from the tetrad and metric of Jordan frame, which are defined as
\begin{equation}\label{trans}
\begin{array}{cccc}
    \tilde g_{\mu\nu}\!\!  &= \Omega^{2} g_{\mu\nu},  \quad &\tilde e\!\! &= \Omega^{4} e,   \\
    \tilde e_{\mu}^{~i}\!\! &= \Omega e_{\mu}^{~i}, \quad &\tilde e^{\mu}_{~i}\!\! &= \Omega^{-1} e^{\mu}_{~i},
\end{array}
\end{equation}
by which one finds that the torsion in Eq. (\ref{torsion}) transforms as
\begin{equation}
    \tilde T^{\rho}_{\mu\nu} =  T^{\rho}_{\mu\nu} + \Omega^{-1} [\delta^{\rho}_{\nu} \partial_{\mu}\Omega - \delta^{\rho}_{\mu} \partial_{\nu}\Omega].
\end{equation}
The torsion scalar transforms as
\begin{equation}
    T = \Omega^2 \tilde T - 4\Omega^{-1}\partial^{\mu}\Omega \tilde T^{\rho}_{~\rho\mu} + 6\Omega^{-2}\partial_{\mu}\Omega \partial^{\mu}\Omega .
\end{equation}
By redefining the scalar field as $\phi\!\!=\!\!\Omega^{2}$
and $\phi\!\!=\!\!e^{\varphi/\sqrt{2\omega-3}}$,
where $\omega \sim 500£¤$ for the observation of the solar system,
and $U(\varphi)\!\!=\!\!\frac{V(\phi)}{\phi^2}$,
the action (\ref{SBD}) can be transformed to the Einstein frame as
constraint
\begin{eqnarray}\label{SE}
    S_{E} &=& \int {\rm d}^4 x~\tilde{e}\,\bigg[\tilde{T} - \frac{2}{\sqrt{2\omega-3}}\tilde \partial^{\mu}\varphi \tilde T^{\rho}_{~\rho\mu} - \frac{1}{2}\tilde g^{\mu\nu}\tilde \nabla_{\mu}\varphi\tilde \nabla_{\nu}\varphi \nonumber \\
    &-& U(\varphi)\bigg] + 2k^2 \int {\rm d}^4 x~\tilde{e}\,\mathcal{\tilde L}_m(\tilde e_{\mu}^{~i}).
\end{eqnarray}
Differing from the case in $f(R)$ gravity,
an additional scalar-torsion coupling term is present in the action.
Therefore, $f(T)$ gravity is not simply dynamically equivalent to
the TEGR action plus a scalar field via conformal transformation,
and one cannot use the results of scalar-tensor theories directly to $f(T)$ gravity.
We investigate the affect of additional scalar-torsion coupling term to the validity
of Birkhoff's theorem in $f(T)$ gravity in our work \cite{fT_birkhoff1,fT_birkhoff2},
and we also analyze the equivalence between the Einstein and the Jordan frames.

In order to obtain the field equation,
we can vary the action (\ref{SE})
with respect to the tetrad field $e_{\alpha}^{~i}$, which yields
\begin{eqnarray}\label{field eqn0}
    4\tilde G^{\alpha}_{~i}
    &=& \frac{2\tilde{e}}{\sqrt{2\omega-3}}\tilde\partial_{\mu}\bigg[\tilde\partial^{\mu}\varphi\frac{\tilde\partial\tilde T^{\rho}_{~\rho\mu}}{\tilde\partial(\tilde\partial_{\mu}\tilde e_{\alpha}^{~i})}\bigg] - \frac{2\tilde\partial^{\mu}\varphi}{\sqrt{2\omega-3}}\frac{\tilde\partial(\tilde{e}\tilde T^{\rho}_{~\rho\mu})}{\tilde\partial\tilde e_{\alpha}^{~i}} \nonumber \\
    &-& \frac{\tilde\partial(\tilde{e}\tilde g^{\mu\nu})}{2\tilde\partial\tilde e_{\alpha}^{~i}}\tilde\nabla_{\mu}\varphi\tilde \nabla_{\nu}\varphi - \frac{\tilde\partial \tilde{e}}{\tilde\partial\tilde e_{\alpha}^{~i}}U(\varphi) + 2k^2\frac{\delta(\tilde{e}\mathcal{\tilde L}_m)}{\delta\tilde e_{\alpha}^{~i}}.
\end{eqnarray}
The left term of above equation is defined as
\begin{equation}
    \tilde G^{\alpha}_{~i} = \tilde\partial_{\mu}(\tilde{e}\tilde e^{\rho}_{~i}\tilde S_{\rho}^{~\mu\alpha}) + \tilde{e}\tilde e^{\nu}_{~i}\tilde T^{\rho}_{~\mu\nu}\tilde S_{\rho}^{~\mu\alpha} - \frac{1}{4}\tilde{e}\tilde e^{\alpha}_{~i}\tilde T.
\end{equation}
According equations (\ref{A1}) and (\ref{A3})
of the additional scalar-torsion coupling term varied
to $e_{\alpha}^{~i}$ and $\partial_{\mu}e_{\alpha}^{~i}$ deduced in the Appendix,
the field equation(\ref{field eqn0}) changes as
\begin{eqnarray}\label{field eqn1}
    \tilde e^{-1} \tilde G^{\alpha}_{~i} &=& \frac{1}{2\sqrt{2\omega-3}}\tilde\partial_{\mu}\big[\tilde\partial^{\alpha}\varphi \tilde e^{\mu}_{~i} - \tilde\partial^{\mu}\varphi \tilde e^{\alpha}_{~i}\big] \nonumber \\
    &-& \frac{\tilde\partial^{\mu}\varphi}{2\sqrt{2\omega-3}}\tilde e^{\alpha}_{~i}\tilde T^{\rho}_{~\rho\mu} +  \frac{\tilde\partial^{\mu}\varphi}{2\sqrt{2\omega-3}}\tilde e^{\rho}_{~i}\tilde T^{\alpha}_{~\rho\mu} \nonumber \\
    &+& \frac{1}{4}\tilde e^{\nu}_{~i}\tilde \nabla^{\alpha}\varphi\tilde \nabla_{\nu}\varphi - \frac{1}{8}\tilde e^{\alpha}_{~i}\tilde \nabla^{\sigma}\varphi\tilde \nabla_{\sigma}\varphi \nonumber \\
    &-& \frac{1}{4}e^{\alpha}_{~i}U(\varphi) + \frac{k^2}{2}e^{\rho}_{~i}\tilde T^{\alpha \,(m)}_{~\rho}.
\end{eqnarray}

With respect to the scalar field $\varphi$, the field equation is obtained by varying the action(\ref{SE}), which yields
\begin{eqnarray}\label{field eqn2}
    -2k^2\frac{\delta(\tilde{e}\,\mathcal{\tilde L}_m)}{\tilde{e}\delta\varphi} &=& \tilde \Box \varphi - \frac{{\rm d}U(\varphi)}{{\rm d}\varphi} \nonumber \\
    &+& \frac{2}{\sqrt{2\omega-3}}\tilde e^{-1}\tilde\partial_{\mu}\big(\tilde{e}\tilde g^{\mu\nu}\tilde T^{\rho}_{~\rho\nu}\big).
\end{eqnarray}

\newcounter{TempEqCnt2}
\setcounter{TempEqCnt2}{\value{equation}}
\setcounter{equation}{20}
\begin{figure*}[ht]
\dotfill
\begin{equation}\label{general_tetrad}
e^i_{~\mu} =
\left( \begin{array}{cccc}
e^{\frac{a(t,r)}{2}} & 0                                                     & 0                         & 0     \\
0                    & {\rm e}^{\frac{b(t,r)}{2}}\!\sin\!\theta \cos\!\phi\, & -r(\cos\!\theta \cos\!\phi \sin\!\gamma + \sin\!\phi \cos\!\gamma)\, & r\!\sin\!\theta(\sin\!\phi \sin\!\gamma - \cos\!\theta \cos\!\phi \cos\!\gamma)  \\
0                    & {\rm e}^{\frac{b(t,r)}{2}}\!\sin\!\theta \sin\!\phi\, & r(\cos\!\theta \cos\!\gamma - \sin\!\phi \sin\!\gamma)\, & -r\!\sin\!\theta(\cos\!\theta \sin\!\phi \cos\!\gamma + \cos\!\phi \sin\!\gamma) \\
0                    & {\rm e}^{\frac{b(t,r)}{2}}\!\cos\!\theta\,            & r\!\sin\!\theta \sin\!\gamma\,           & r\!\sin^2\!\theta \cos\!\gamma . \\
\end{array} \right).
\end{equation}
\dotfill
\end{figure*}
\setcounter{equation}{\value{TempEqCnt2}}
\setcounter{equation}{21}

\section{The Validity of Birkhoff's Theorem for $f(T)$ Gravity via Perturbative Approach}

The basic of $f(T)$ gravity is vierbein field $\mathbf{e}_{i}(x^{\mu})$ and Weitzenb\"ock connection.
This theory is not invariant under local Lorentz transformations,
so different tetrads will lead to different results.
Using a local Lorentz transformation in the tangent space,
people can construct general tetrad for the spherically symmetric metric \cite{fT10},
which is shown as Eq.(\ref{general_tetrad}) between the dotted lines at the top of the page.
And $\gamma$ is the new degree of freedom of the $f(T)$ theory
due to the lack of local Lorentz invariance \cite{fT_Lorentz}.
If we set $\gamma\!\!=\!\!-\pi/2$, this tetrad field reduces to the off diagonal tetrad
considered in our previous work \cite{fT_birkhoff2}.
Using this general tetrad field, via the tensor operation (\ref{basic}), we will get the spherically symmetric metric written in the following form with arbitrary values of $\theta$, $\phi$ and $\gamma$:
\begin{equation}\label{metric}
    {\rm d} s^2={\rm e}^{a\!(r,t)}~{\mathrm{d}}t^2-{\rm e}^{b\!(r,t)}~{\mathrm{d}}r^2-r^2{\mathrm{d}}\theta^2-r^2 \sin^2\!\theta~{\mathrm{d}}\phi^2.
\end{equation}

It is well known that for Einstein's field equations, the only solution in vacuum for a spherically symmetric metric is given by the Schwarzschild solution, or Schwarzschild-(A)dS solution if a cosmological constant is included in the field equations. This result, called Birkhoff's theorem, was proved independently by G. D. Birkhoff \cite{birkhoff_B} and J. T Jebsen \cite{birkhoff_J}.
Here, we investigate the validity of Birkhoff's theorem for $f(T)$ gravity via perturbative approach, which means one should perform perturbations around a spherically symmetric solution. Therefore, we can analyze the physically equivalence between the Einstein and the Jordan frames.
In the perturbation forms, the tetrad and scalar fields can be written as
\begin{eqnarray}
    \tilde e_\alpha^{~i} &=& \tilde e_\alpha^{~i(0)} + \tilde e_\alpha^{~i(1)}, \nonumber \\
    \varphi  &=& \varphi^{(0)} + \varphi^{(1)},
\end{eqnarray}

We have investigate the result of perturbation using this general tetrad field.
At zero order, it is the same as the result with diagonal tetrad field.
Birkhoff's theorem is valid and the solution is the well known
Schwarzschild-(A)dS metric.
But the higher-order perturbation is too complex and unable to process.
Therefore, we will use the case of the diagonal tetrad field as a representative at zero-order perturbation,
and analyze in detail the case of the diagonal tetrad field at the higher order perturbation.

\subsection{zero-order perturbation}

The case of the diagonal tetrad field is more easy to express than that of the general tetrad field.
Correspondingly, the spherically symmetric tetrad field can be written in the following diagonal form:
\begin{displaymath}
\left\{ \begin{array}{lll}
\tilde e_t^{~0}      & = & e^{\frac{a(r,t)}{2}}\approx e^{\frac{a^{(0)}(r,t)}{2}}e^{\frac{a^{(1)}(r,t)}{2}} \\
\tilde e_r^{~1}      & = & e^{\frac{b(r,t)}{2}}\approx e^{\frac{b^{(0)}(r,t)}{2}}e^{\frac{b^{(1)}(r,t)}{2}} \\
\tilde e_\theta^{~2} & = & r \\
\tilde e_\psi^{~3}   & = & r \sin\theta
\end{array} \right.
\end{displaymath}
while the inverse tetrad field, satisfying the relation $e^{\alpha}_{~i}\cdot e_{\beta}^{~i}\!\!=\!\!\delta^{\alpha}_{\beta}$, can be given by
\begin{displaymath}
\left\{ \begin{array}{lll}
\tilde e^t_{~0}      & = & e^{\frac{-a(r,t)}{2}}\approx e^{\frac{-a^{(0)}(r,t)}{2}}e^{\frac{-a^{(1)}(r,t)}{2}} \\
\tilde e^r_{~1}      & = & e^{\frac{-b(r,t)}{2}}\approx e^{\frac{-b^{(0)}(r,t)}{2}}e^{\frac{-b^{(1)}(r,t)}{2}} \\
\tilde e^\theta_{~2} & = & \frac{1}{r} \\
\tilde e^\psi_{~3}   & = & \frac{1}{r \sin\theta}
\end{array} \right.
\end{displaymath}

The perturbations also act on the scalar potential $U(\varphi)$,
such that the potential can be expanded around a background solution for the scalar field $\varphi_{0}$
\begin{equation}
    U(\varphi)=\sum_{n} \frac{U^{n(0)}}{n!}\big(\varphi-\varphi^{(0)} \big)^{n}.
\end{equation}

By inserting the above expressions into the field equations (\ref{field eqn1}) and (\ref{field eqn2}),
we can split the equations into the different orders of perturbations.
We are interested in vacuum solutions where $\tilde T^{\alpha \,(m)}_{~\rho}\!\!=\!\!0$,
and we consider the background solution of the scalar field to be a constant at zero-order $\varphi^{(0)}(r,t)\!\!=\!\!\varphi_{0}$, for the reason that
at zero-order the results for TEGR framework has to be recovered.
Equations (\ref{field eqn1}) and (\ref{field eqn2})
at zero order are given by
\begin{eqnarray}
    \!\!0 \!\!&=&\!\! \tilde e^{-1} \tilde G^{\alpha(0)}_{~i} + \frac{1}{2}\tilde e^{\alpha}_{i}\Lambda , \nonumber \\
    \!\!0 \!\!&=&\!\! \tilde e^{-1} \tilde\partial_{\mu}(\tilde{e}\tilde e^{\rho}_{~i}\tilde S_{\rho}^{~\mu\alpha}) + \tilde e^{\nu}_{~i}\tilde T^{\rho}_{~\mu\nu}\tilde S_{\rho}^{~\mu\alpha} - \frac{1}{4}\tilde e^{\alpha}_{~i}\tilde T + \frac{1}{2}\tilde e^{\alpha}_{i}\Lambda , \label{zero order}
\end{eqnarray}
and
\begin{eqnarray}
    \frac{{\rm d}U(\varphi^{(0)})}{{\rm d}\varphi} &=& \frac{2}{\sqrt{2\omega-3}}\tilde e^{-1}\tilde\partial_{\mu}\big(\tilde{e}\tilde g^{\mu\nu}\tilde T^{\rho}_{~\rho\nu}\big).
\end{eqnarray}
Here the cosmological constant is defined as $U_{0}\!\!=\!\!2\Lambda$.
Equation (\ref{zero order}) is the Einstein-like equation in the TEGR framework
with a cosmological constant. For convenience, we introduce the tensor $E^\mu_{~i}$ to represent of
the right hand side of Eq. (\ref{zero order}), then the field
equation can be re-expressed as
\begin{equation}
    E^\mu_{~i}= 0.
\end{equation}
Then we work out all the components of $E^\mu_{~i}$, and find nearly half of them are not vanishing, including some quite complicated ones. Three of which we used, fortunately
not very complex, are given by, respectively
\begin{eqnarray}
E^{r}_{~0}  &=& \frac{\dot b^{(0)}(r,t)}{2e^{b^{(0)}(r,t)}r} \label{E10}\\
E^{t}_{~0}  &=& \frac{b^{(0)}(r,t)^\prime r-1+e^{b^{(0)}(r,t)}+\Lambda e^{b^{(0)}(r,t)}r^{2}}{2e^{b^{(0)}(r,t)}r^{2}} \label{E00}\\
E^{r}_{~1}  &=& \frac{-a^{(0)}(r,t)^\prime r-1+e^{b^{(0)}(r,t)}+\Lambda e^{b^{(0)}(r,t)}r^{2}}{2e^{b^{(0)}(r,t)}r^{2}}.  \label{E11}
\end{eqnarray}
These three terms are the same as the results in the case of the general tetrad field,
so the solution is general for any tetrad field.
For the perfect fluid models of matter, the non-diagonal elements of energy-momentum tensor are naturally equal
to zero, which limits $E^{r}_{~0}$ to be zero. Eq. (\ref{E10}) restricts $b^{(0)}(r,t)$ to be only a function of $r$, that is
\begin{equation}\label{B}
b^{(0)}(r,t)=b^{(0)}(r).
\end{equation}
Contrasting Eq. (\ref{E00}) with Eq. (\ref{E11}), leads to the result that
\begin{equation}\label{constrain}
a^{(0)}(r,t)^\prime = -b^{(0)}(r)^\prime.
\end{equation}
For $b^{(0)}$ is independent of $t$, the left of Eq. (\ref{constrain}) should be also a function of $r$. As long as the solution exists, the function $a^{(0)}(r,t)$ could be simply expressed as
\begin{equation}\label{A}
a^{(0)}(r,t)=\widetilde a^{(0)}(r)+c(t),
\end{equation}
where $c(t)$ is an arbitrary function of $t$.
Therefore the function ${\rm e}^{a(r,t)}$ can be written as
\begin{equation}
{\rm e}^{a^{(0)}(r,t)}={\rm e}^{\widetilde a^{(0)}(r)}{\rm e}^{c(t)}.
\end{equation}
The factor ${\rm e}^{c(t)}$ can always be absorbed
in the metric through a coordinate transformation $t\to t^\prime$,
where $t^\prime$ is a new time coordinate defined as
\begin{equation}
 {\rm d}t^\prime={\rm e}^{\frac{c(t)}{2}} {\rm d}t.
\end{equation}
After solving the Eq. (\ref{E00}),
the solution is the well known Schwarzschild-(A)dS metric,
which gives the zero-order solution as
\begin{equation}
    e^{a^{(0)}(r)} = e^{-b^{(0)}(r)} = 1 - \frac{2 m}{r} - \frac{\Lambda}{3}r^2.
\end{equation}
where $m$ is an integration constant.

Then, at zero order we have a static metric which satisfies Birkhoff's theorem given above,
and this result is valid, no matter how to choose any kind of tetrad field.

\subsection{first-order perturbation}

At the first linear order, as we have said before,
the result of the general tetrad field is too complex and unable to process.
Therefore here, we can only analyze in detail the case of the diagonal tetrad field for higher perturbation.

Due to $ \partial_{\mu}\varphi^{(0)}\!\!=\!\!0$,
the field equations (\ref{field eqn1}) and (\ref{field eqn2}) for the first-order perturbation are simplified as
\begin{eqnarray}
    \tilde e^{-1} \tilde G^{\alpha(1)}_{~i} &=& \frac{1}{2\sqrt{2\omega-3}}\tilde\partial_{\mu}\big[\tilde\partial^{\alpha}\varphi^{(1)} \tilde e^{\mu(0)}_{~i} - \tilde\partial^{\mu}\varphi^{(1)} \tilde e^{\alpha(0)}_{~i}\big] \nonumber \\
    &-& \frac{\tilde\partial^{\mu}\varphi^{(1)}}{2\sqrt{2\omega-3}} \tilde e^{\alpha(0)}_{~i}\tilde T^{\rho(0)}_{~\rho\mu} + \frac{\tilde\partial^{\mu}\varphi^{(1)}}{2\sqrt{2\omega-3}} \tilde e^{\rho(0)}_{~i}\tilde T^{\alpha(0)}_{~\rho\mu}  \nonumber \\
    &-& \frac{1}{4}\tilde e^{\alpha(1)}_{~i}U_{0}(\varphi_{0}) - \frac{1}{4}\tilde e^{\alpha(0)}_{~i}U_{0}'(\varphi_{0}) \varphi^{(1)}  \label{first order}
\end{eqnarray}
and
\begin{eqnarray}
    U_{0}''(\varphi) \varphi^{(1)} &=& \tilde \Box \varphi^{(1)} + \frac{2}{\sqrt{2\omega-3}}\tilde e^{-1}\tilde\partial_{\mu}\big(\tilde{e}\tilde g^{\mu\nu}\tilde T^{\rho(1)}_{~\rho\nu}\big).
\end{eqnarray}
And the definite form of $\tilde G^{\alpha(1)}_{~i}$ is
\begin{eqnarray}
    \tilde G^{\alpha(1)}_{~i}
    &=& \tilde\partial_{\mu}\bigg[\tilde{e}\tilde e^{\rho(1)}_{~i}\tilde S_{\rho}^{~\mu\alpha(0)} + \tilde{e}\tilde e^{\rho(0)}_{~i}\tilde S_{\rho}^{~\mu\alpha(1)}\bigg]  \nonumber \\
    &-& \frac{1}{4}\tilde{e}\tilde e^{\alpha(1)}_{~i}\tilde T^{(0)} - \frac{1}{4}\tilde{e}\tilde e^{\alpha(0)}_{~i}\tilde T^{(1)} + \tilde{e}\tilde e^{\nu(1)}_{~i}\tilde T^{\rho(0)}_{~\mu\nu}\tilde S_{\rho}^{~\mu\alpha(0)} \nonumber \\
    &+& \tilde{e}\tilde e^{\nu(0)}_{~i}\tilde T^{\rho(1)}_{~\mu\nu}\tilde S_{\rho}^{~\mu\alpha(0)} + \tilde{e}\tilde e^{\nu(0)}_{~i}\tilde T^{\rho(0)}_{~\mu\nu}\tilde S_{\rho}^{~\mu\alpha(1)}.
\end{eqnarray}
In the above calculation, the deduce of $\tilde T^{\rho}_{\mu\nu}$ is according to the redefined scalar field as $\phi\!\!=\!\!\Omega^{2}$, $\phi\!\!=\!\!e^{\varphi/\sqrt{2\omega-3}}$ and $\varphi^{(0)}(r,t)\!\!=\!\!\varphi_{0}$, so we can simplify them as
\begin{eqnarray}
    \tilde T^{\rho}_{\mu\nu} &=& T^{\rho}_{\mu\nu} + \frac{1}{2\sqrt{2\omega-3}}(\delta^{\rho}_{\nu} \partial_{\mu}\varphi - \delta^{\rho}_{\mu} \partial_{\nu}\varphi) , \\
    \tilde T^{\rho(0)}_{\mu\nu} &=& T^{\rho(0)}_{\mu\nu}.
\end{eqnarray}
We can get some special torsion tensor solutions by using the diagonal tetrad field for the subsequent calculation
\begin{eqnarray}
    \tilde T^{t(0)}_{~rt} &=& \frac{1}{2}a^{(0)}(r)' + \frac{\tilde\partial^{r}\varphi^{(0)}}{2\sqrt{2\omega-3}} = \frac{1}{2}a^{(0)}(r)'  , \\
    \tilde T^{r(0)}_{~tr} &=& \frac{1}{2} \dot b^{(0)}(r) + \frac{\dot \varphi^{(0)}}{2\sqrt{2\omega-3}} = 0.
\end{eqnarray}

The first linear order field equations look quite terrible for calculation,
therefore we only consider the following three special components of Eq. (\ref{first order}):
\begin{eqnarray}
    \tilde e^{-1} \tilde G^{t(1)}_{~1}
    \!\!&=&\!\! \frac{1}{2\sqrt{2\omega-3}}\tilde\partial_{r}(\dot \varphi^{(1)} \tilde e^{r(0)}_{~1}) + \frac{\dot \varphi^{(1)}}{2\sqrt{2\omega-3}}\tilde e^{r(0)}_{~1}\tilde T^{t(0)}_{~rt} \nonumber \\
    \!\!&=&\!\! \frac{e^{\frac{1}{2}a^{(0)}(r)}}{2\sqrt{2\omega-3}} \bigg(\tilde\partial_{r} \tilde\partial^{t}\varphi^{(1)} + \dot \varphi^{(1)} a^{(0)}(r)'  \bigg) ,
    \\
    \tilde e^{-1} \tilde G^{r(1)}_{~0}
    \!\!&=&\!\! \frac{1}{2\sqrt{2\omega-3}}\tilde\partial_{t}(\tilde\partial^{r}\varphi^{(1)} \tilde e^{t(0)}_{~0}) + \frac{\tilde\partial^{r}\varphi^{(1)}}{2\sqrt{2\omega-3}}\tilde e^{t(0)}_{~0}\tilde T^{r(0)}_{~t r} \nonumber \\
    \!\!&=&\!\! \frac{e^{-\frac{1}{2}a^{(0)}(r)}}{2\sqrt{2\omega-3}} \tilde\partial_{t} \tilde\partial^{r}\varphi^{(1)} ,
    \\
    \tilde e^{-1} \tilde G^{t(1)}_{~2} \!\!&=&\!\! 0.
\end{eqnarray}
Importing the definite form of $\tilde G^{\alpha(1)}_{~i}$, the first linear order field equations give that
\begin{equation}\label{F01}
    \frac{\dot a^{(1)}(r,t) e^{-\frac{1}{2}a^{(1)}(r,t)}}{2 r e^{-a^{(0)}(r)}}
    = \frac{e^{\frac{1}{2}a^{(0)}(r)}}{2\sqrt{2\omega-3}}\bigg[\tilde\partial_{r} \tilde\partial^{t}\varphi^{(1)} + \dot \varphi^{(1)}a^{(0)}(r)' \bigg] ,
\end{equation}
\begin{equation}\label{F10}
    -\frac{\dot b^{(1)}(r,t) e^{\frac{1}{2}b^{(1)}(r,t)}}{2 r}
    = \frac{e^{-\frac{1}{2}a^{(0)}(r)}}{2\sqrt{2\omega-3}} \tilde\partial_{t} \tilde\partial^{r}\varphi^{(1)} ,
\end{equation}
\begin{equation}\label{F02}
    -\frac{\cos\theta\bigg(\dot a^{(1)}(r,t) e^{-\frac{1}{2}a^{(1)}(r,t)} + \dot b^{(1)}(r,t) e^{\frac{1}{2}b^{(1)}(r,t)}\bigg)}{4 r^2\sin\theta} = 0.
\end{equation}
From Eq. (\ref{F02}), we can find the relation
\begin{equation}
    \dot a^{(1)}(r,t) e^{-\frac{1}{2}a^{(1)}(r,t)} = -\dot b^{(1)}(r,t) e^{\frac{1}{2}b^{(1)}(r,t)}.
\end{equation}
Considering this relation to Eq. (\ref{F10}),
and contrasting Eq. (\ref{F10}) with Eq. (\ref{F01}),
it is easy to find an important constraint $\dot \varphi^{(1)}\!\!=\!\!0$,
which makes Eq. (\ref{F10}) and Eq. (\ref{F01}) to change as
\begin{eqnarray}
    \frac{\dot a^{(1)}(r,t) e^{-\frac{1}{2}a^{(1)}(r,t)}}{2 r e^{-a^{(0)}(r)}} &=& 0 , \\
    -\frac{\dot b^{(1)}(r,t) e^{\frac{1}{2}b^{(1)}(r,t)}}{2 r} &=& 0.
\end{eqnarray}
From the two above equations, one can obviously find that $\dot a^{(1)}(r,t)\!\!=\!\!0$ and $\dot b^{(1)}(r,t)\!\!=\!\!0$.
So considering the first-order perturbation, Birkhoff's theorem still holds in the Einstein frame.
Because of $\varphi^{(0)}\!\!=\!\!\varphi_{0}\!\!=\!\!const$ and $\varphi^{(1)}\!\!=\!\!\varphi(r)$,
which is consistent with the analysis for diagonal tetrad in our previous work \cite{fT_birkhoff2}.
So the conformal transformation relation does not depend on time.
Then we transform back the metric from Einstein frame to Jordan frame,
consequently, the metric in the Jordan frame clearly does not depend on time,
indicating that Birkhoff's theorem is still satisfied in first-order perturbation.
In the situation of the higher order, the violation of Birkhoff's theorem may appear,
which will respond to the non-physically equivalence between the Einstein frame and the Jordan frame.

Calculating other components of Eq. (\ref{first order}),
leads to the result that $\tilde G^{\theta(1)}_{~2}\!\!=\!\!\tilde G^{\psi(1)}_{~3}\sin\!\theta$,
which yields
\begin{equation}
    \frac{(\sin^2\!\theta - 1)}{2} \bigg(e^{a^{(0)}(r)}a^{(0)}(r)' + e^{a^{(1)}(r)}a^{(1)}(r)'\bigg) = 0.
\end{equation}
Consequently, we find that
\begin{eqnarray}
    e^{a^{(1)}(r)}a^{(1)}(r)' &=& -e^{a^{(0)}(r)}a^{(0)}(r)'  \nonumber \\
    e^{a^{(1)}(r)}  &=& W_{1} - e^{a^{(0)}(r)} \nonumber \\
    &=& W_{1}-1 + \frac{2 m}{r} + \frac{\Lambda}{3}r^2 ,
\end{eqnarray}
where $W_{1}$ is a constant. One other useful component is $\tilde G^{r(1)}_{~2}\!\!=\!\!0$, which yields
\begin{eqnarray}
    0 &=& \frac{\cot\!\theta}{4 r^2} \bigg(b^{(1)}(r)' e^{-\frac{1}{2} b^{(1)}(r)} + a^{(0)}(r)' e^{-\frac{1}{2}b^{(1)}(r)}\bigg) , \nonumber \\
    b^{(1)}(r) &=& W_{2} - a^{(0)}(r) , \nonumber \\
    e^{b^{(1)}(r)}  &=& \frac{e^{W_{2}}}{\bigg( 1 - \frac{2 m}{r} - \frac{\Lambda}{3}r^2 \bigg)}.
\end{eqnarray}
where $W_{2}$ is also a constant.
Considering that space-time background tends to be flat where $r\!\!\rightarrow\!\!\infty$,
which means that $e^{a^{(0)}(r)}\!\!\rightarrow\!\!1$ or $a^{(0)}(r)\!\!\rightarrow\!\!0$,
the perturbation terms $a^{(1)}(r)$ and $b^{(1)}(r)$ naturally tend to zero.
Therefore, we can assume $W_{1}\!\!=\!\!2$ and $W_{2}\!\!=\!\!0$, which yields
\begin{eqnarray}
    e^{a(r)} &=& e^{a^{(0)}(r)}e^{a^{(1)}(r)} \nonumber \\
    &=& \bigg( 1 - \frac{2 m}{r} - \frac{\Lambda}{3}r^2 \bigg) \bigg( 1 + \frac{2 m}{r} + \frac{\Lambda}{3}r^2\bigg)  \\
    e^{b(r)} &=& e^{b^{(0)}(r)}e^{b^{(1)}(r)} \nonumber \\
    &=& \bigg( 1 - \frac{2 m}{r} - \frac{\Lambda}{3}r^2 \bigg)^{-2}.
\end{eqnarray}
Through the Newton approximation,
under the definition of $g_{\mu\nu}\!\!=\!\!\eta_{\mu\nu}+h_{\mu\nu}$,
we finally get $g_{0 0}\!\!=\!\!1 + 2 V$,
where we have defined $c\!\!=\!\!1$ and the Newton potential $V$.
If we take the zero-order Schwarzschild solution,
which does not consider the cosmological constant effect,
The Newton law of gravitation can be deduced.
Considering the first-order perturbation solution,
which is different with Schwarzschild solution,
therefore, we can examine this precise solution
or determine the solution parameters up to first order
through fitting the cosmological data-sets,
such as galaxy rotation curve or Pioneer anomaly.

\section{Discussions and Conclusions}

In our previous work,
we investigated Birkhoff's theorem with diagonal tetrad field \cite{fT_birkhoff1}
and the extended Birkhoff's theorem with off diagonal tetrad field \cite{fT_birkhoff2}.
Here, we continue to investigate the range of validity of Birkhoff's theorem
with the general tetrad field for $f(T)$ gravity by using a perturbative approach.
Assuming a constant scalar field as the background solution,
we can see that the zero-order solution in perturbations gives a static metric,
but the higher-order perturbation is too complex and unable to process.
So we can only analyze in detail the case of the diagonal tetrad field at higher-order perturbation.
And the first linear order solution provides a tetrad field that is
time-independent in the Einstein frame via conformal transformation,
leading that Birkhoff's theorem is hold.
Parallelly, we find that the result obtained in the Einstein frame
on the range of validity of Birkhoff's theorem
is not affected when one returns to the Jordan frame
for the time-independent constraint on the $\varphi$ field,
where the tetrad field is also static at first order in perturbations.
Hence, this can show to a certain degree the physical equivalence
between the Jordan and the Einstein frames at least in perturbation order.
In the situation with the higher-order perturbation, the violation of Birkhoff's theorem may appear,
which will respond to the non-physically equivalence between the both frames.
This result is not obviously contradictory to that of $f(R)$ theory \cite{fR_perturbation},
which cannot constrain the time-independent relation of the $\varphi$ field at first-order perturbation.
If the time-independent constraint on the $\varphi$ field also existed in $f(R)$ theory,
Birkhoff's theorem with the diagonal tetrad field would still hold
in the Jordan frame up to the first linear order perturbation.
This difference is very similar to the discussion of our previous work \cite{fT_birkhoff2}.
The extra six degrees of freedom in the off diagonal tetrad or the general tetrad
conceal the physical meaning of the (time-dependent) $\varphi$ field.
When we choose the specially diagonal tetrad field solution for $f(T)$ gravity,
some additional constraints are introduced,
exactly as the time-independent relation of the $\varphi$ field.

The classical Birkhoff's theorem not only gives the
unique solution to the spherically symmetric distribution gravity source,
but also sheds lights on the gravity collapse phenomena.
Up to the first-order perturbation as shown, Birkhoff's theorem may still hold in some cases,
and the first-order perturbation solution has been obtained.
Therefore one can apply these results to study the
gravity collapsing phenomena via perturbative approach.
We will continue the related gravity collapsing research.

\section*{Acknowledgements}

This work is partly supported by National Natural
Science Foundation of China under Grant Nos. 11075078 and 10675062
and by the project of knowledge Innovation Program (PKIP) of Chinese
Academy of Sciences (CAS) under the grant No. KJCX2.YW.W10 through
the KITPC where we have initiated this present work.
Xin He MENG would also like to thank Prof. Lewis H.Ryder and Prof. Sergei D.Odintsov for helpful discussions,
and to dedicate this work to remember one of his teachers once working in Tucson of USA,
who has inspired him for exploring the truth, beauty, fairness and justice of this world forever!

\section{Appendix}

Here, we deduce in detail the variation of the action
for a simply scalar torsion $T$ with respect to vierbein.
Considering the basic relation
$e^\alpha_{~i}\,e_\beta^{~i}\!\!=\!\!\delta ^\alpha_{~\beta}$ in teleparallel gravity,
we can define the algebraic complement $C^\alpha_{~i}$ of $e_\alpha^{~i}$,
\begin{eqnarray}
  e &=& det(e_\alpha^{~i})
     =  e_\alpha^{~i} \, C^\alpha_{~i} \\
  \frac{\delta e}{\delta e_\alpha^{~i}} &=& C^\alpha_{~i}.
\end{eqnarray}
So we get $C^\alpha_{~i}\!\!=\!\!e\,e^\alpha_{~i}$£¬and the variation of metric with respect to $e_\alpha^{~i}$
\begin{eqnarray}
    g^{\mu\nu} &=& \eta^{ij} e^{\mu}_{~i} e^{\nu}_{~j}  \nonumber \\
    \frac{\delta g^{\mu\nu}}{\delta e_\alpha^{~i}} &=& \frac{\eta^{jk} \delta (e^{\mu}_{~j} e^{\nu}_{~k})}{\delta e_\alpha^{~i}}
                                                    =  -2 g^{\mu\alpha}e^{\nu}_{~i}.
\end{eqnarray}
and
\begin{eqnarray}
    g_{\mu\nu} &=& \eta_{ij} e_{\mu}^{~i} e_{\nu}^{~j}  \nonumber  \\
    \frac{\delta g_{\mu\nu}}{\delta e_\alpha^{~i}} &=& \frac{\eta_{jk} \delta (e_{\mu}^{~j} e_{\nu}^{~k})}{\delta e_\alpha^{~i}}
                                                    =  2 g_{\mu\alpha} e^{\nu}_{~i}.
\end{eqnarray}
Redefining the energy-momentum tensor formula of $e_\alpha^{~i}$
and $e\!\!=\!\!\sqrt{-g}$
\begin{eqnarray}
  T_{\mu\nu} &=& -\frac{2}{\sqrt{-g}}\frac{\delta(\sqrt{-g} \mathcal{L}_m)}{\delta g^{\mu\nu}} \nonumber\\
             &=&  \frac{1}{e}\frac{\delta (e \mathcal{L}_m)}{\delta e_\alpha^{~i}}\frac{1}{g^{\mu\alpha}e^{\nu}_{~i}}.
\end{eqnarray}
Maybe a more suitable form for this work is
\begin{eqnarray}
  e^{\rho}_{~i} T^\alpha_{~\rho} &=& \frac{1}{e}\frac{\delta (e \mathcal{L}_m)}{\delta e_\alpha^{~i}} .
\end{eqnarray}
Differing from that in Einstein's theory of general relativity,
the teleparallel gravity uses Weitzenb\"ock connection, defined directly from
the vierbein (\ref{vierbein}) and antisymmetric non-vanishing torsion (\ref{torsion}).

Then we can deduce the variation of the torsion tensor with respect to $e_\alpha^{~i}$
and $\partial_{\mu}e_{\alpha}^{~i}$, respectively,
\begin{eqnarray}
    \frac{\delta T^{\rho}_{~\mu\alpha}}{\delta e_\alpha^{~i}}
&=& \frac{\delta[e^{\rho}_{~i}(\partial_{\mu}e_{\alpha}^{~i} - \partial_{\alpha}e_{\mu}^{~i})]}{\delta e_\alpha^{~i}} \nonumber \\
&=& - e^\alpha_{~i}T^{\rho}_{~\mu\alpha} , \label{A1}\\
    \frac{\delta T^{\rho}_{~\mu\alpha}}{\delta(\partial_{\mu}e_{\alpha}^{~i})}
&=& \frac{\delta[e^{\rho}_{~i}(\partial_{\mu}e_{\alpha}^{~i} -
    \partial_{\alpha}e_{\mu}^{~i})]}{\delta(\partial_{\mu}e_{\alpha}^{~i})} \nonumber \\
&=& 2 e^{\rho}_{~i} , \label{A2}
\end{eqnarray}
and the coupling with $\tilde\partial^{\mu}\varphi$
\begin{eqnarray}
\frac{\tilde\partial^{\mu}\varphi \delta T^{\rho}_{~\rho\mu}}{\delta(\tilde\partial_{\mu}\tilde e_{\alpha}^{~i})}
&=& \tilde\partial^{\mu}\varphi(\delta^{\mu}_{\rho}\tilde e^{\rho}_{~i})\delta^{\alpha}_{\mu} - \tilde\partial^{\mu}\varphi(\tilde e^{\rho}_{~i}\delta^{\alpha}_{\rho}) \nonumber \\
&=& \tilde\partial^{\alpha}\varphi \tilde e^{\mu}_{~i} - \tilde\partial^{\mu}\varphi \tilde e^{\alpha}_{~i}. \label{A3}
\end{eqnarray}

The other two tensors are defined by (\ref{contorsion}) and (\ref{S}).
Then the torsion scalar as the teleparallel Lagrangian is defined by (\ref{T}).
One needs to pay attention to the $S_{\rho}^{~\mu\nu}$ being 
a polynomial combination of the product of $g^{\mu\nu}$ and $T^{\rho}_{~\mu\nu}$, like
\begin{equation}
    S_\rho^{~\mu\nu} = \sum g^{a b} \cdot T^{c}_{~d e}.
\end{equation}
The indices $a,b,c,d,e$ of the above definition are dummy indices.
After summation of these five dummy indices, only $\rho,\mu,\nu$ are left.
Then we can use a step-by-step method for the binomial formula to
deduce the variation of the torsion scalar with respect to $e_\alpha^{~i}$
and $\partial_{\mu}e_{\alpha}^{~i}$, respectively,
\begin{eqnarray}
    \frac{\delta T}{\delta e_{\alpha}^{~i}}
&=& \frac{\delta S_{\rho}^{~\mu\nu}}{\delta e_{\alpha}^{~i}}T^{\rho}_{~\mu\nu}
 +  S_{\rho}^{~\mu\nu} \frac{\delta T^{\rho}_{~\mu\nu}}{\delta e_{\alpha}^{~i}} \nonumber \\
&=& \left( \frac{\delta S_{\rho}^{~\mu\nu}}{g^{\mu\nu}} \frac{g^{\mu\nu}}{\delta e_{\alpha}^{~i}} + \frac{\delta S_{\rho}^{~\mu\nu}}{T^{\rho}_{~\mu\nu}} \frac{T^{\rho}_{~\mu\nu}}{\delta e_{\alpha}^{~i}} \right)
T^{\rho}_{~\mu\nu} + S_{\rho}^{~\mu\nu} \frac{\delta T^{\rho}_{~\mu\nu}}{\delta e_{\alpha}^{~i}} \nonumber \\
&=& -4 e^{\beta}_{~i} T^{\rho}_{~\mu\beta} S_\rho^{~\mu\alpha} ,
\end{eqnarray}
and
\begin{eqnarray}
    \frac{\delta T}{\delta (\partial_{\mu}e_{\alpha}^{~i})}
&=& \frac{\delta S_{\rho}^{~\mu\nu}}{\delta (\partial_{\mu}e_{\alpha}^{~i})}T^{\rho}_{~\mu\nu}
 +  S_{\rho}^{~\mu\nu} \frac{\delta T^{\rho}_{~\mu\nu}}{\delta (\partial_{\mu}e_{\alpha}^{~i})} \nonumber \\
&=& \frac{\delta S_{\rho}^{~\mu\nu}}{\delta T^{\rho}_{~\mu\nu}} \frac{\delta T^{\rho}_{~\mu\nu}}{\delta (\partial_{\mu}e_{\alpha}^{~i})} \cdot T^{\rho}_{~\mu\nu} + S_{\rho}^{~\mu\nu} \frac{\delta T^{\rho}_{~\mu\nu}}{\delta (\partial_{\mu}e_{\alpha}^{~i})} \nonumber \\
&=& 4 e^{\rho}_{~i} S_{\rho}^{~\mu\alpha}.
\end{eqnarray}
Finally, we can get the variation equation (\ref{field eqn})
of the action (\ref{action}) with respect to the vierbein.

%

%
\end{document}